\documentclass[aps,prl,twocolumn,superscriptaddress,showpacs]{revtex4-1}
\usepackage{ucs}
\usepackage{graphicx}
\usepackage{setspace}
\usepackage{units}
\usepackage{xspace}
\usepackage{color}
\usepackage[normalem]{ulem}
\usepackage{textcomp}
\usepackage{units}
\usepackage{siunitx}
\usepackage{amsmath}
\usepackage{amsfonts}
\usepackage{amssymb}

\newcommand{\blue}[1]{\textcolor{black}{#1}}
\usepackage{soul} 

\DeclareSIUnit\muB{$\mu_B$}

\begin{document}

\title{Instability of the topological surface state in Bi$_2$Se$_3$ upon deposition of gold}
 \author{A. Polyakov}
 \affiliation{Max-Planck-Institut f\"ur Mikrostrukturphysik, Weinberg 2, 06120 Halle, Germany}
 \author{C. Tusche}
 \affiliation{Peter Gr\"unberg Institut (PGI-6), Forschungszentrum J\"ulich GmbH, 52425 J\"ulich, Germany}
 \affiliation{Max-Planck-Institut f\"ur Mikrostrukturphysik, Weinberg 2, 06120 Halle, Germany}
 \author{M. Ellguth}
 \affiliation{Johannes-Gutenberg-Universit\"at Mainz, Institut f\"ur Physik, Staudingerweg 7, 55116 Mainz, Germany}
 \affiliation{Max-Planck-Institut f\"ur Mikrostrukturphysik, Weinberg 2, 06120 Halle, Germany}
 \author{E.\,D. Crozier}
 \affiliation{Department of Physics, Simon Fraser University Burnaby, BC Canada, V5A 1S6}
 \author{K. Mohseni}
 \affiliation{Max-Planck-Institut f\"ur Mikrostrukturphysik, Weinberg 2, 06120 Halle, Germany}
 \author{M.\,M. Otrokov}
 \affiliation{Donostia International Physics Center (DIPC), 20018 San Sebasti\'an/Donostia, Basque Country, Spain}
 \affiliation{Tomsk State University, 634050 Tomsk, Russia}
 \affiliation{Saint Petersburg State University, 198505 Saint Petersburg, Russia}
 \author{X. Zubizarreta}
 \affiliation{Max-Planck-Institut f\"ur Mikrostrukturphysik, Weinberg 2, 06120 Halle, Germany}
 \author{M.\,G.~Vergniory}
 \affiliation{Max-Planck-Institut f\"ur Mikrostrukturphysik, Weinberg 2, 06120 Halle, Germany}
 \affiliation{Donostia International Physics Center (DIPC), 20018 San Sebasti\'an/Donostia, Basque Country, Spain}
 \author{M. Geilhufe}
 \affiliation{Max-Planck-Institut f\"ur Mikrostrukturphysik, Weinberg 2, 06120 Halle, Germany}
 \affiliation{Nordita, Center for quantum materials, KTH Royal Institute of Technology and Stockholm University, Roslagstullsbacken 23, SE-106 91 Stockholm, Sweden}
 \author{E.\,V. Chulkov}
 \affiliation{Donostia International Physics Center (DIPC), 20018 San Sebasti\'an/Donostia, Basque Country, Spain}
 \affiliation{Tomsk State University, 634050 Tomsk, Russia}
 \affiliation{Departamento de F\'{\i}sica de Materiales UPV/EHU,Centro de F\'{\i}sica de Materiales CFM - MPC and Centro Mixto CSIC-UPV/EHU, 20080 San Sebasti\'an/Donostia, Basque Country,Spain}
 \author{A. Ernst}
 \affiliation{Max-Planck-Institut f\"ur Mikrostrukturphysik, Weinberg 2, 06120 Halle, Germany}
 \affiliation{Institut f\"ur Theoretische Physik, Johannes Kepler Universit\"at, A 4040 Linz, Austria}
 \email{theory: aernst@mpi-halle.de}
\author{ H.\,L. Meyerheim}
 \affiliation{Max-Planck-Institut f\"ur Mikrostrukturphysik, Weinberg 2, 06120 Halle, Germany}
 \email{experiment: hmeyerhm@mpi-halle.de}
 \author{S.\,S.\,P. Parkin}
 \affiliation{Max-Planck-Institut f\"ur Mikrostrukturphysik, Weinberg 2, 06120 Halle, Germany}
\date{\today}

\begin{abstract}
  Momentum resolved photoemission spectroscopy indicates the
  instability of the Dirac surface state upon deposition of gold on
  the (0001) surface of the topological insulator
  Bi$_{2}$Se$_{3}$. Based on the structure model derived from extended
  x-ray absorption fine structure experiments showing that gold atoms
  substitute bismuth atoms, first principles calculations provide
  evidence that a gap appears due to hybridization of the surface
  state with gold $d$-states near the Fermi level. Our findings
  provide new insights into the mechanisms affecting the stability of
  the surface state.
\end{abstract}
\pacs{61.05.cp, 73.20.At, 71.15.Mb, 79.60.-i}
\maketitle

Topological protection of the Dirac electrons at the three-dimensional
topological insulator (TI) surface caused enormous interest in these
materials as potential candidates for spintronics
\cite{Hasan2010,Wehling2014}. This remarkable property guarantees that
the Dirac electrons of a TI surface do not experience the
backscattering whatever the surface quality is. The only crucial
condition to be met to secure that behavior is the absence of the
time-reversal symmetry breaking perturbations, for instance, an
out-of-plane ferromagnetism. Therefore, since the very discovery of
the three-dimensional TIs, considerable experimental effort has been
devoted to the confirmation of the topological protection. The most
striking evidences of the property include the absence of the elastic
backscattering on the disordered or defected surfaces
\cite{Roushan2009, Hanaguri.prb2010, Alpichshev2010, Sessi.prb2013},
existence of the well-defined Dirac states at the thallium-based 3D
TI's surfaces exhibiting complex morphology \cite{Kuroda.prl2010a,
  Kuroda.prb2013}, and the tolerance of the topological states to the
in-plane~\cite{Honolka2012,Scholz2012} or, possibly,
non-collinear~\cite{Polyakov2015} magnetic moments.

The property of the topological protection is intimately related to
the topological surface state (TSS) integrity. In particular, the
time-reversal symmetry leads to a crossing of the surface states at
the 2D Brillouin zone center.  To gap out the Dirac point (DP) an
effective mass term has to be taken into account, e.g. by applying a
magnetic field in $z$
direction~\cite{Chen2010,Vergniory2014,Wray2011,Henk2012,Otrokov.prb2015,Sessi2016}.
Otherwise, both experimentally and theoretically, the DP and TSS were
found to be robust upon deposition of various adsorbates
\cite{Zhu2011, Benia.prl2011, King2011, Roy2014} and overlayers
\cite{Miao.pnas2013, Menshchikova.nl2013, Shokri.prb2015}. However
recently, it has been found that the TSS can be destroyed by strain in
the vicinity of grain boundaries on the surface of epitaxial
Bi$_2$Se$_3$(0001) thin films~\cite{Liu2014} and in
Pb$_{1-x}$Sn$_x$Te~\cite{Geilhufe2015a}.  More recently, for the bulk
alloy (Bi$_{1-x}$Mn$_x$)$_2$Se$_{3}$, the influence of Mn-induced
ferromagnetic order was excluded from being responsible for the formation
of the 100~meV band gap~\cite{Sanchez-Barriga2016}. It was argued that
the system remains topologically non-trivial while in-gap resonance
states of d-symmetry are involved in the gap
opening~\cite{Sanchez-Barriga2016}.  This view is supported by
theoretical studies suggesting the formation of impurity (vacancy)
resonance states near the DP~\cite{Biswas2010,Black-Schaffer2012,
  Black-Schaffer2012a}, located in the bulk or in the near surface
region. These resonance states hybridize with the TSS, which is
destroyed and the DP is energetically split. As pointed out in
Ref.~\onlinecite{Black-Schaffer2012a} the topological protection of
the TSS is only valid for two-dimensional backscattering but there is
no protection against scattering by bulk states, which may originate
from non-magnetic and magnetic impurities.

Since the experiment in Ref.~\cite{Sanchez-Barriga2016} only dealt
with bulk alloys it remains an open question whether the disruption of
the TSS also occurs in the case of a surface alloy prepared by in situ
deposition of (sub-) monolayer amounts of an adsorbate on a TI
surface. \blue{As was already proven before, a topological surface
  state is generated by the electronic structure of the bulk and is
  mainly located in the two first quintuple layers
  (QL)~\cite{Hasan2010}. Therefore, one cannot exclude that the TSS in
  the experiment in Ref.~\cite{Sanchez-Barriga2016} is destroyed due
  to substantial changes in structural or electronic properties of the
  whole sample. In addition, the magnetic nature of the band gap
  opening was not fully ruled out: Mn can form clusters with nonzero
  magnetization. Also, each Mn atom possesses a magnetic moment, which
  can strongly interact with the free electron gas. Such an
  electron-magnon interaction can also induce the band gap
  opening~\cite{Chotorlishvili2014}.}

\blue{In order to elucidate the mechanism by which the TSS is modified
  by doping with an impurity we have carried out a combined
  experimental and theoretical study to investigate the effect of a
  prototype \emph{non-magnetic}} surface alloy like gold on the (0001)
surface of a Bi$_{2}$Se$_{3}$ \blue{single crystal}, in the following
written as (Bi$_{1-x}$Au$_{x}$)$_{2}$Se$_{3}$. The structure analysis
by extended x-ray absorption fine structure (EXAFS) experiments
provides evidence that gold atoms deposited at T=160~K substrate
temperature substitute bismuth atoms within the near-surface
regime. Simultaneously, photoemission experiments \blue{carried out at
  0.3, 0.5 and 1~ML coverage} indicate the \blue{opening of a gap at
  the DP} of the TSS \blue{even at a coverage of 0.3~ML equivalent to
  x=0.15.} Here, and in the following we refer to 1~ML as
6.74$\times$10$^{14}$ atoms/cm$^{2}$, i.e., one ad-atom per surface
atom.  \textit{Ab-initio} calculations based on the model structure
show that the \blue{modification of the TSS} already begins between
x=0.0125 and x=0.0375. They perfectly reproduce the experimental
spectral function recorded by photoemission. We conclude that impurity
states of $d$-symmetry located within the topmost QL are responsible
for the gap opening. Our results provide an experimental and
theoretical proof that even \blue{nonmagnetic} surface impurities
create resonance states near the DP, involving scattering by bulk like
states~\cite{Biswas2010,Black-Schaffer2012,Black-Schaffer2012a}.

EXAFS measurements were carried out at the Sector 20 insertion device
beamline at the Advanced Photon Source (APS), Argonne National
Laboratory (USA) using the MBE1 end station equipped with standard
surface analytical tools~\cite{Gordon2007}. Two samples were
investigated corresponding to a film coverage of 0.3 and
0.6~monolayers (ML) of gold as estimated by AES spectra in combination
with reflection high energy electron diffraction oscillations.  After
gold deposition on the pristine Bi$_{2}$Se$_{3}$(0001) surface EXAFS
data were collected above the Au-L$_{III}$ absorption edge
(E$_{0}$=11919~eV) in the fluorescent yield (FY) mode using a
4-element Vortex Si drift detector. From AES experiments it was
concluded that in-diffusion of gold into the Bi$_{2}$Se$_{3}$ bulk
sets in above 250~K. For this reason gold deposition and EXAFS
measurements were carried out at T=160~K. Monochromatic x-rays from a
Si(111) double-crystal monochromator, with the 7 GeV APS ring
operating in top-up mode, were incident on the substrate at
approximately 2/3 of the critical angle
($\alpha_{c}\approx 0.4^{\circ}$) for total reflection to avoid errors
due to anomalous dispersion effects~\cite{Jiang1998}. The in-plane and
out-of-plane gold atomic environment was investigated exploiting the
polarization dependence of the linearly polarized x-ray beam with the
electric field vector aligned either perpendicular (E$_{\perp}$) or
parallel (E$_{\parallel}$) to the plane of the substrate.

\begin{figure}[t]
  \center
  \includegraphics[width=0.85\columnwidth]{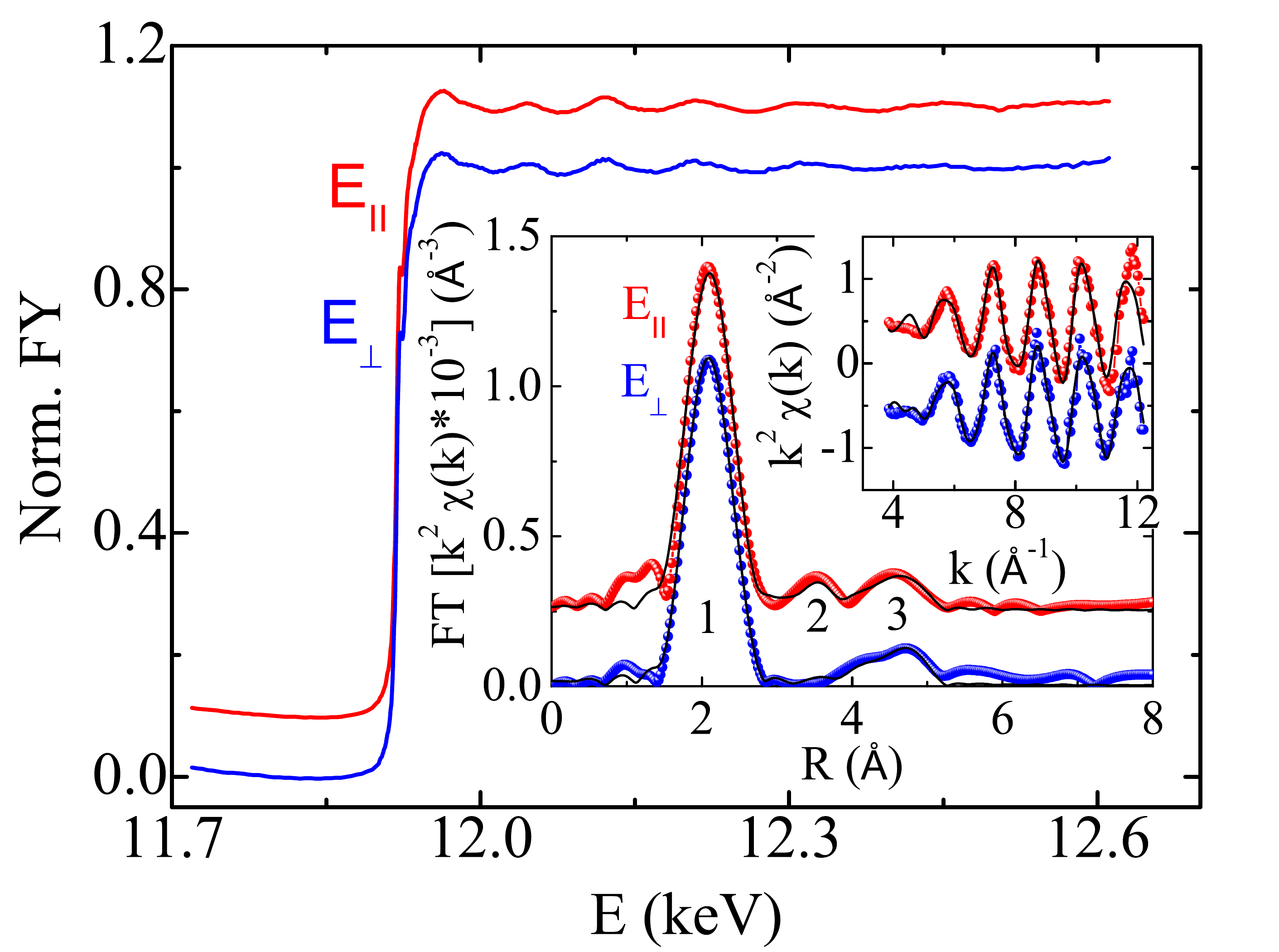}
  \caption{(Color online) Normalized fluorescence yield collected for
   about 0.6 ML of gold deposited on Bi$_{2}$Se$_{3}$(0001) with the
   electric field vector parallel (E$_{\parallel}$) and perpendicular
   (E$_{\perp}$) to the sample surface. Curves are shifted for
   clarity. The inset shows the k$^{2}$-weighted
   interference function [$\chi(k)$] and the magnitude of the
   Fourier-Transforms. The absence of any polarization dependence
   of peak 1 indicates a nearly isotropic environment compatible with the
   substitutional site (see Fig.~\ref{Fig2}). The reduction of peak 2 in the
   magnitude of the Fourier Transform of E$_{\perp}$ is predicted for the
   substitutional site.} \label{Fig1}
\end{figure}

Fig.~\ref{Fig1} shows for the 0.6 ML sample the fluorescent yield (FY)
versus photon energy for both geometries which is proportional to the
linear absorption coefficient. The inspection of the figure reveals
that both curves are nearly identical. The detailed analysis was
carried out by background subtraction and calculation of the
Fourier-Transform (FT) of the k$^{2}$-weighted interference function
[$\chi(k)$], which are shown in the inset.

The FT magnitude shows three peaks labelled by 1, 2 and 3 for
E$_{\parallel}$, but only 1 and 3 for E$_{\perp}$.  Peak 1 corresponds
to the first selenium shell, while peaks 2 and 3 are related to
bismuth shells. For the present discussion we focus on the first
shell, which was fitted in R-space using theoretical scattering
amplitudes and phases. The fit results are listed in Table~\ref{Tab}.  In
detail, we find that the first peak corresponds to in total six
selenium atoms of which three are located at a distance of 2.45~\AA~
and another three at a distance of 2.65~\AA~(E$_{\parallel}$). For
E$_{\perp}$ geometry the distances can be viewed as identical. The structure
parameters are compatible with a model, in which gold atoms substitute
bismuth atoms. 

Comparison of the experimental first shell distances with those in the
un-relaxed bulk (R=2.87~\AA~ and R=3.07~\AA) indicates substantial local
relaxations of the surrounding selenium atoms upon bismuth
substitution. This situation closely resembles that of iron deposition
on Bi$_{2}$Se$_{3}$(0001)~\cite{Polyakov2015}, where very similar
structure parameters have been derived.

\begin{figure}[t]
  \center
  \includegraphics[width=0.80\columnwidth]{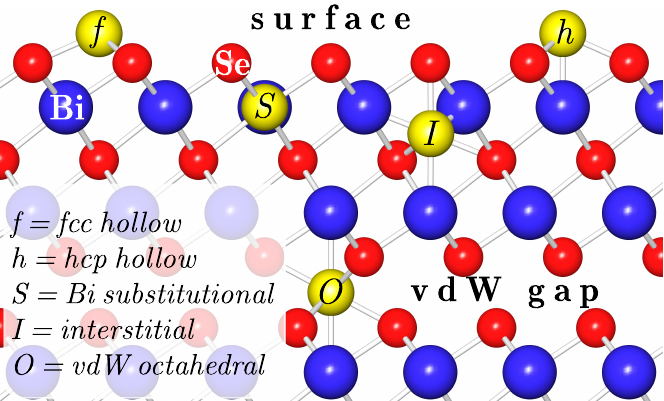}
  \caption{(Color online) Schematic view of adsorption sites for gold
    on Bi$_{2}$Se$_{3}$(0001). Bismuth and selenium atoms are shown as
    large blue and small red spheres, respectively. Possible
    gold positions are indicated by yellow spheres and labelled
    by "f", "h", "S", "I", and "O" for the fcc, hcp, substitutional,
    interstitial and octahedral van der Waals gap sites, respectively.} \label{Fig2}
\end{figure}

The structure model is discussed on the basis of Fig.~\ref{Fig2},
where the near surface structure of Bi$_{2}$Se$_{3}$ including the
first QL is schematically shown. Large (blue) and small (red) spheres
represent bismuth and selenium atoms, respectively. The
Bi$_{2}$Se$_{3}$ crystal structure is characterized by an
ABCBA stack of van-der-Waals (vdW) bonded QLs each being composed
of a Se-Bi-Se-Bi-Se sequence of layers~\cite{Roy.prb2014}. Possible
adsorption positions for gold are indicated and are labelled as "fcc"
(f), "hcp" (h), "substitutional" (S), "interstitial"(I) and
"octahedral gap" (O). The absence of any polarization dependence of
the peak 1 as well as its relation to a distance in the 2.5 to
2.7~\AA~range leads to a model where gold atoms are substituting
bismuth atoms. Other adsorption sites such as the "fcc" or "hcp" sites
at the surface or in interstitial sites are not compatible with both
the isotropy of the first peak and its relation to a Au-Se bond
length.

\begin{table}
\label{Tab}
\caption{Table of structural parameters for 0.6~ML Au on
  Bi$_{2}$Se$_{3}$. The meaning of the parameters is as follows: 
  R:=~refined neighbor distance, R$_{b}$:=~distance in unrelaxed bulk
  structure, N$^{*}$:=~effective polarization dependent 
  coordination number for L$_{III}$ edge \cite{Citrin1985},
  $\sigma^{2}$:=~mean squared relative displacement  amplitude,
  $\Delta$E$_{0}$:~shift of absorption edge, R$_{u}$:=~Residual in
  percent~\cite{RUEXAFS}. The amplitude reduction factor (S$_{0}^{2}$)
  was kept constant at S$_{0}^{2}$=0.80 in all cases. Parameters
  labelled by an asterisk (*) are kept fixed.
  Uncertainties are given in brackets.}

\begin{center}
 \begin{tabular}{l|c|c|c|c|c|c|c|c}

 Pol. & Sh & R$_{b}$(\AA) & N & R (\AA) & N$^{*}$ &
 $\sigma^{2}$(\AA$^{2}$)&$\Delta$ E$_{0}$ (eV)& R$_{u}$ \\[0.5ex]

 \hline\hline

       \toprule

        E$\parallel$  & Se1 & 2.87 & 3 &  2.45 (3) & 3.04(*) & 0.004 & 2.6&  4.9  \\
                      & Se2 & 3.07 & 3 &  2.65 (4) & 2.92(*) & 0.014 & 2.6&       \\

 \hline\hline

      E$\perp$        & Se1 & 2.87 & 3 &  2.44 (3) & 2.92(*) & 0.004 & 3.3&  2.3  \\
                      & Se2 & 3.07 & 3 &  2.66 (4) & 3.16(*) & 0.021 & 3.3&       \\

 \end{tabular}

\end{center}
\end{table}

\begin{figure*}[htb]
  \center{\includegraphics[width=.8\textwidth]{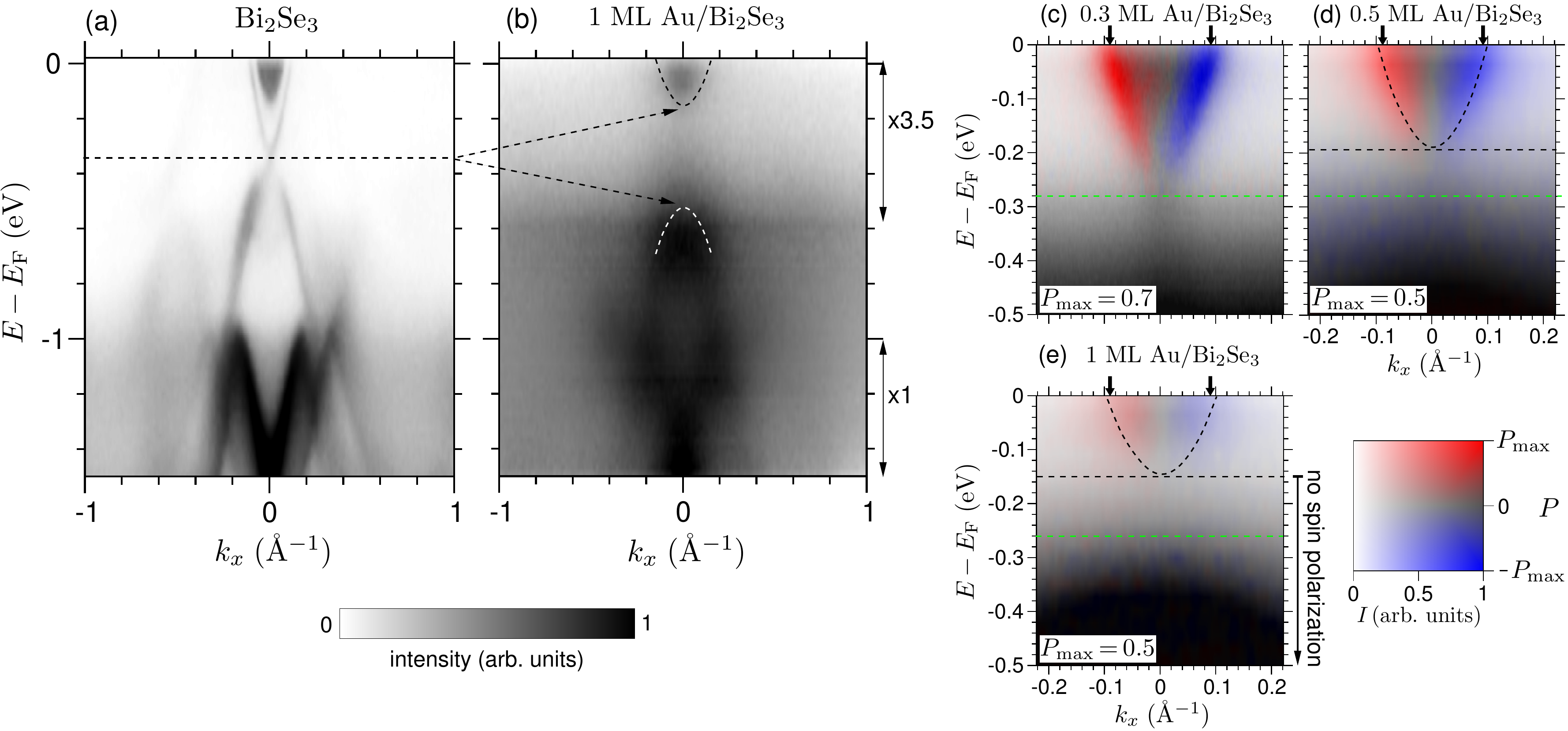}}
  \caption{\blue{(a-b): Spin integrated photoemission spectra of
      pristine and gold-covered Bi$_2$Se$_3$ using unpolarized He-I
      radiation ($h\nu\,{=}\,\SI{21.2}{eV}$). (c-e): Spin-resolved
      spectra using s-polarized laser light
      ($h\nu\,{=}\,\SI{6.0}{eV}$) collected for 0.3, 0.5 and 1.0~ML of
      gold. Dashed lines indicate the opening of a gap in the surface
      state. Further details are provided in the Supplementary Materials.}}
  \label{Fig3}
\end{figure*}

\blue{Spin-resolved photoemission experiments were carried out using a
  momentum microscope (MM)~\cite{Tusche2015} equipped with a 2D
  imaging spin filter based on low-energy electron diffraction at 1 ML
  Au/Ir(001)~\cite{Vasilyev2015}. Sample preparation which closely
  followed the procedure employed for the EXAFS experiments and
  experimental details are reported in the Supplementary
  Materials. The photoemission spectra collected by the MM for
  pristine and gold-covered Bi$_2$Se$_3$ (0001) are displayed along
  the $\overline{\mathrm{M}}-\overline{\Gamma}-\overline{\mathrm{M}'}$
  direction in Fig.~\ref{Fig3}(a-e), respectively.  In the case of the
  pristine Bi$_2$Se$_3$ sample [Fig.~\ref{Fig3}(a)], the DP is located
  approximately 350~meV below $E_\text{F}$ resulting from $n$ doping
  by selenium vacancies.  After deposition of 1~ML of gold
  [Fig.~\ref{Fig3}(b)], the TSS is strongly broadened by the
  structural disorder introduced by the randomly distributed gold
  atoms on Bi sites.}

\blue{More details are derived by spin-resolved spectra collected for
  0.3, 0.5 and 1~ML as shown in Figs. 3(b-e). Here, the
  characteristic spin-momentum locking of the surface state is
  observed. Gold deposition induces an upwards shift
  (p-doping) and -more importantly- an opening of the gap which
  amounts to about 100~meV. The surface state dispersion is not
  observed to continue into the energy range below the black dashed
  lines where the two spin-polarized branches meet at $k_x\,{=}\,0$
  [see Fig.~\ref{Fig3}(d,e)]. The experimental findings are well
  reproduced by ab-initio calculations as discussed in the following.}

 \begin{figure*}[t]
 \center
 \includegraphics[width=0.8\textwidth]{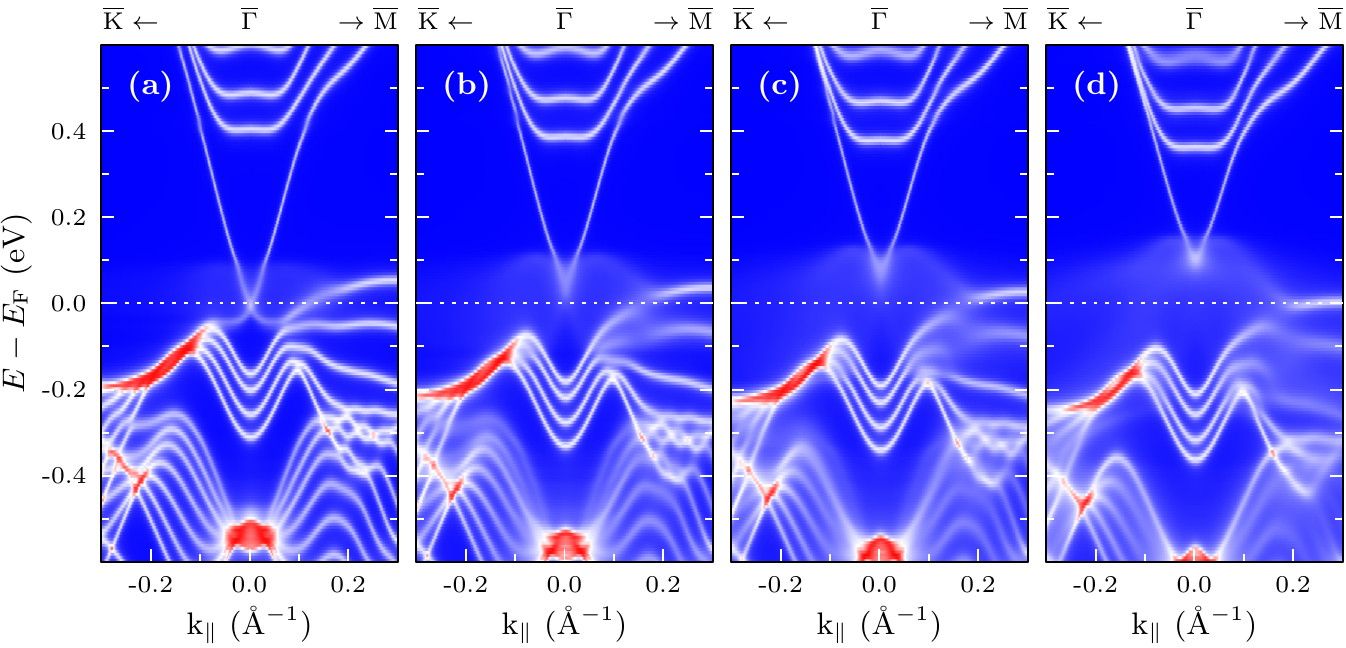}
 \caption{(Color online) Evolution of the TSS in
   (Bi$_{1-x}$Au$_x$)$_2$Se$_{3}$/Bi$_2$Se$_3$(0001): Calculated spectral
   density for $x=0.0125$ (a), $x=0.0375$ (b), $x=0.0625$ (c), and
   $x=0.1250$ (d), respectively.}
 \label{Fig4}
 \end{figure*}

 \blue{First-principles calculations of Au/Bi$_2$Se$_3$ (0001) model
  gold induced modifications of the TSS.} We
 used a self-consistent full relativistic Green function method
 especially designed for semi-infinite materials such as surfaces and
 interfaces~\cite{Luders2001, Geilhufe2015}. Alloying the bismuth layers
 with gold was simulated within a coherent potential approximation as
 implemented within the multiple scattering
 theory~\cite{Soven1967,Gyorffy1972}. The structural information was
 adopted from the EXAFS experiments (see Table~\ref{Tab}).

 In order to trace the evolution of the TSS upon gold concentration 
 the $k$-resolved spectral density was calculated in
 (Bi$_{1-x}$Au$_{x}$)$_2$Se$_{3}$ on the Bi$_2$Se$_3$(0001) surface
 for $\overline{\mathrm{K}}-\overline{\Gamma}-\overline{\mathrm{M}}$
 direction in SBZ at concentrations $x=\{0.0125,0.0375,0.0625,0.125\}$
 (see (Fig.~\ref{Fig4}(a-d), respectively). In all these cases Au is
 assumed to be homogeneously distributed over the Bi sites within the first QL. 
 The results are shown in Fig.~\ref{Fig4}, where low, medium and high spectral density
 is represented by blue, white and red color coding, respectively. \blue{At $x=0.0125$
 [see Fig.~\ref{Fig4}(a)], the spectral density is similar to that of
 the pristine Bi$_{2}$Se$_{3}$~(0001) surface (not shown), but some broadening is seen in the
 vicinity of E$_{F}$, which is attributed to a gold resonance
 state.}

The most important result is that gold substitution induces the
formation of a gap of $\Delta$E=200 meV, in a good agreement
with photoemission spectra.  \blue{Within the gap regime} the spectral density is
considerably weakened due to the presence of a resonance state and
because of band broadening induced by structural disorder.
Furthermore, gold acts as a p-dopant shifting the bands up in
energy. Comparison of the calculated band structure of the pristine
surface (not shown here) and the alloyed sample indicate an upward
shift of the bands by approximately 330 meV (at $x=0.5$).

Our study provides new insight into the non-magnetic impurity mediated
modification of the TSS. Gold atoms deposited in the sub- to one
monolayer range on Bi$_{2}$Se$_{3}$(0001) occupy substitutional
bismuth sites as evidenced by EXAFS measurements. This goes in
parallel with the dramatic weakening of the spectral density of the
TSS, thus opening a gap as observed by photoemission.  In accordance
with first-principles calculations, gold in bismuth substitutional
site within the first QL creates a d-type resonant state near the
$E_{F}$, which strongly hybridizes with the bands of the TI and
substantially modifies its surface electronic structure.  The surface
alloy involving only the topmost QL is sufficient for the gap opening,
which we attribute to the fact that the resonance state near E$_{F}$
is of $d$-symmetry.  According to the model of Black-Schaffer and
Balatsky~\cite{Biswas2010,Black-Schaffer2012, Black-Schaffer2012a} a
bulk-surface interaction is a prerequisite for the opening of the gap,
since the TSS is not protected by scattering processes involving bulk
three-dimensional states.

 We acknowledge financial support from DFG through priority program
 SPP1666 (Topological Insulators), the Spanish Ministry of Science
 and Innovation (Grants No. FIS2013-48286-C02-02-P and No.
 FIS2013-48286-C02-01-P), the Basque Government through
 the Nanomaterials project under the nanoGUNE2014 program
 (Grant No. IE05-151), the Tomsk State University Academic
 D. I. Mendeleev Fund Program in 2015 (Research Grant No. 8.1.05.2015),
 and partial support from Saint Petersburg State University
 (Project No. 15.61.202.2015). Technical support by F. Weiss is
 gratefully acknowledged. We also thank Z.S. Aliev, M. B. Babanly,
 K. A. Kokh, and O. E. Tereshchenko for support
 by providing samples. M.E. acknowledges support from the BMBF (Grant
 No.~05K12EF1). E.D.C. acknowledges research grants from the
 Dean, Faculty of Science, Simon Fraser University, and from
 NSERC. Sector 20 facilities at the Advanced Photon Source, and
 research at these facilities, are supported by the US Department of
 Energy - Basic Energy Sciences, the Canadian Light Source and its
 funding partners, and the Advanced Photon Source. Use of the Advanced
 Photon Source, an Office of Science User Facility operated for the
 U.S. Department of Energy (DOE) Office of Science by Argonne National
 Laboratory, was supported by the U.S. DOE under Contract
 No. DE-AC02-06CH11357. M.G. acknowledges European Research Council
 under the European Union’s Seventh Framework Program (FP/2207-2013)/ERC
 Grant Agreement No. DM-321031.
 \bibliographystyle{apsrev}
 \bibliography{./AuBi2Se3}

\begin{thebibliography}{42}
\bibitem[{RUE()}]{RUEXAFS}
\bibinfo{note}{The unweighted residuum (Ru) is defined by:
  R$_u=\sum|(Yi_{obs}-Y_{calc})|/\sum(Y_{obs})$, where Y$_{obs}$ and Y$_{calc}$
  are the experimental and calculated magnitudes of the peak selected in the FT
  of the $\chi(k)$ spectrum and the summation extends over all points in the
  window used to filter the peak.}
\end{thebibliography}
 
\clearpage
\onecolumngrid
\begin{appendix}
\section{Supplementary material}
\subsection{Photoemission experiment}
Gold was deposited at T=223~K, followed by a mild annealing at 243~K
for 5 minutes, the latter improving the structural order, but without
inducing in-diffusion of gold in the bulk as in the main text.

For the spin ntegrated photoemission measurements, the
Au/Bi$_{2}$Se$_3$ (0001) sample was illuminated by He-I radiation
(21.2 eV) under an angle of incidence of 22$^\circ$ with respect to
the sample surface, and the azimuth along the
$\overline{\mathrm{K}}-\overline{\Gamma}-\overline{\mathrm{K}'}$
direction in the surface Brillouin zone (SBZ). Photoelectrons emitted
into the complete solid angle above the sample surface (emission
angles of $\pm90^\circ$) were collected by the objective lens of a
photoelectron momentum microscope. The image of the photoemission
intensity as function of the lateral crystal momentum (k$_x$,k$_y$) is
obtained in the focal plane of the objective lens not requiring any
sample movement. Series of constant energy momentum images in steps of
10\,meV were recorded by a CCD camera from a fluorescent screen image
detector.For the spin-resolved photoemission experiments we introduce
a 2D imaging spin filter based on low-energy electron diffraction at 1
ML Au/Ir(001) into the momentum microscope.  For the spin resolved
measurements shown in Fig.~3c-d (main text), the sample was
illuminated by 6.0~eV s-polarized laser radiation from the 4th
harmonic of a Ti:Sa oscillator, incident along the
$\overline{\mathrm{M}}-\overline{\Gamma}-\overline{\mathrm{M}'}$
direction of the SBZ.

\begin{figure}[htb]
   \center{\includegraphics[width=0.95\columnwidth]{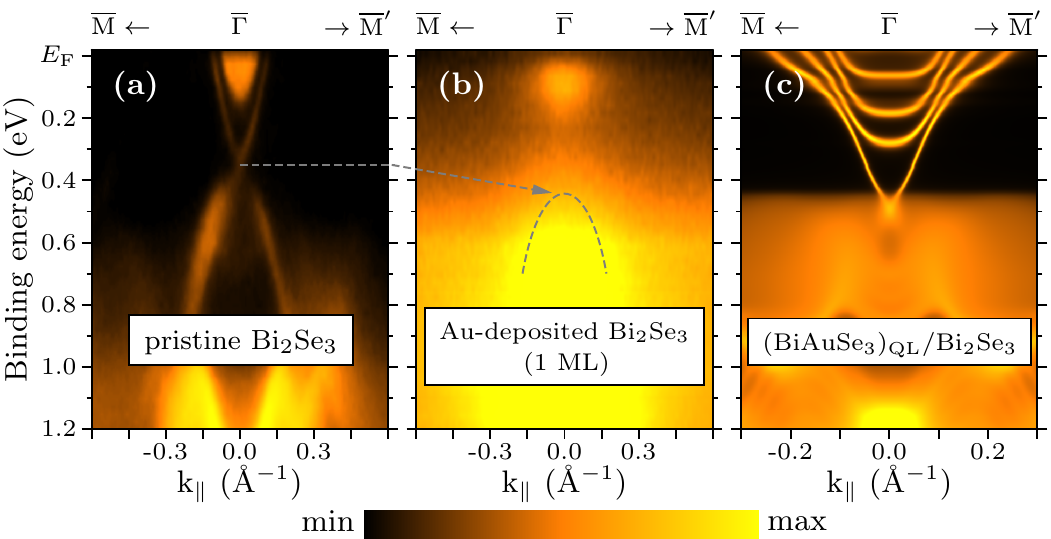}}
   \caption{Photoemission of the pristine Bi$_2$Se$_3$ (a), and after
     deposition of the equivalent of 1 ML gold (b). The spectra,
     obtained by excitation with unpolarized He-I radiation
     ($h\nu\,{=}\,\SI{21.2}{eV}$), are compared with the
     calculated band structure for x=0.35,0.25,0.25,0.15
     for the first four Bi layers, respectively (c).
     The Fermi level in (c) is approximately set to the
     experimental one for a clear comparison.}
   \label{FigS1}
\end{figure}

First, the surface band structure of 1~ML gold substituted into
bismuth sites and distributed over several QL's in the surface
vicinity was calculated. Since the layer-resolved occupancy of Bi
sites by 1ML Au was not known from the experiment, we tried several
configurations comparing the calculated electronic structures with
the photoemission experiment. A good agreement was achieved for the
case of Au distributed over the first two QL's with concentrations
x=0.35,0.25,0.25,0.15 in the first four Bi layers, respectively. The
calculated electronic structure shown in Fig.~\ref{FigS1}(c) agrees
well with the experimental photoemission spectra (see
Fig.~\ref{FigS1}(b)). The most important result is that gold
substitution induces the formation of a gap of $\Delta$E=0.2 eV,
which is in a good agreement with the photoemission experiment.  The
spectral density is considerably weakened due to the presence of a
resonance state and because of band broadening induced by structural
disorder.  Furthermore, gold acts as a p-dopant shifting the bands up
in energy. Comparison of the calculated band structure of the
pristine surface (not shown here) and the alloyed sample indicate an
upward shift of the bands by approximately 0.33 eV (at x=0.5).

The photoemission spectra collected by the MM for pristine and
gold-covered Bi$_2$Se$_3$ (0001) are displayed along the
$\overline{\mathrm{M}}-\overline{\Gamma}-\overline{\mathrm{M}'}$
direction in Fig.~\ref{FigS1}(a,b), respectively. In the case of the
pristine Bi$_2$Se$_3$ sample [Fig.~\ref{FigS1}(a)], the DP is located
approximately 350~meV below $E_\text{F}$ resulting from $n$ doping by
selenium vacancies. After the deposition of gold equivalent to one ML
[Fig.~\ref{FigS1}(b)], the TSS is strongly broadened by the
structural disorder introduced by the randomly distributed gold atoms
on Bi sites, which affects the TSS, as seen in in the calculated
spectrum [Fig.~\ref{FigS1}(c), see the discussion below]. The
photoemission process itself is additionally affected by this
disorder, leading to a higher background intensity. Therefore, the
intensity of the TSS above the DP is too much weakened to be
observed. At lower energies than the DP, however, the broadened TSS
feature is still observed. As indicated schematically in
Figs.~\ref{FigS1}(a,b), the original dispersion of the TSS below the
DP is shifted to lower energies, consistent with the formation of a
gap.

While the photoemission peak near $E_\text{F}$ in Fig.~3(b) (main text)
looks similar to the bulk bands near $E_\text{F}$ in
Fig.~3(a), the spin-resolved spectra for 0.3, 0.5 and 1 ML
Au/Bi$_2$Se$_3$ in Fig.~3(c-e) show the characteristic spin
momentum locking of the surface state (opposite spin polarization for
${\pm}\,k$). Upon the deposition of gold, several effects are
observed. First, the dispersion of the spin polarized surface state is
energetically compressed instead of being shifted [note the small
black arrows in Fig.~3(c-e) indicating that the
$k_x$-diameter at $E_\text{F}$ is not varying as much as would result
from a shift of the surface-state dispersion]. Second, the spin
polarization is conserved after the gold deposition. This means that
the surface state is not destroyed but modified. The surface state
dispersion is not observed to continue in the energy range below the
black dashed lines where the two spin-polarized branches meet at
$k_x\,{=}\,0$ [see Fig.~3(d,e)]. In the same energy range,
the spin polarization vanishes. Contrarily, for a clean Bi$_2$Se$_3$,
the surface state is known to be spin-polarized below the Dirac
point. These observations support our interpretation in terms of a gap
opening.

\subsection{Impurity states in Au/Bi$_2$Se$_3$ (0001)}

\begin{figure}[]
 \center
 \includegraphics[width=0.7\columnwidth]{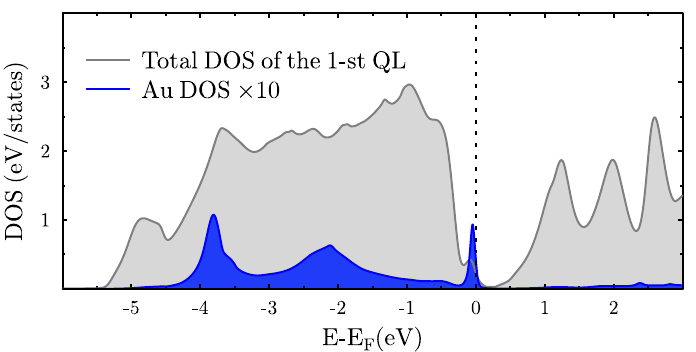}
 \caption{(Color online) Density of states of the first QL in
   Bi$_{1.975}$Au$_{0.025}$Se$_{3}$/Bi$_2$Se$_3$(0001). The blue area
   shows the DOS of gold scaled by factor 10.}
 \label{FigS2}
 \end{figure}

The corresponding density of states (DOS) for x=0.0125 is shown in
Fig.~\ref{FigS2}, where the blue (dark) partial DOS represent the
contribution of gold to the total DOS (bright). The resonance is
clearly seen near E$_{F}$. This resonance state is of $d$-character
and present in (Bi$_{1-x}$Au$_{x}$)$_{2}$Se$_{3}$ throughout the whole
concentration range investigated ($0 < x\le 0.5$). Owing to the
$d$-symmetry of the resonance state it is very effective regarding the
modification of the TSS since it is related to a strong hybridization
with the semiconductor states. With increasing gold concentration the
DP is split off from E$_{F}$ and a gap appears. The size of the gap is
increasing with the gold concentration (x) and approaches the maximum
value of $\Delta$E=150 meV at $x=0.5$.
\end{appendix} 
 
\end{document}